\documentclass[aps,prl,showpacs,amsmath,amssymb,aps,preprintnumbers,twocolumn,superscriptaddress]{revtex4-2}
\usepackage{amsmath,amssymb}
\usepackage{bm}
\usepackage{tipa}
\usepackage{upgreek}
\usepackage{comment}
\usepackage{mathrsfs}
\usepackage{graphicx}
\usepackage{braket}
\usepackage{enumitem}
\usepackage{mathbbol}
\usepackage{booktabs}
\usepackage{gensymb}
\usepackage{physics}
\usepackage[normalem]{ulem}
\usepackage{color}
\usepackage{xcolor}
\usepackage[colorlinks,bookmarks=true,citecolor=blue,linkcolor=red,urlcolor=blue]{hyperref}
\usepackage{hyperref}

\usepackage{pifont}

\usepackage{physics}
\usepackage{bm}

\newcommand{\e}{\mathrm{e}}
\newcommand{\ii}{\mathrm{i}}

\newcommand{\PLLL}{\mathcal P_{\mathrm{LLL}}}

\begin{document}

\author{Haydn S. Adlong}
\affiliation{Institute for Theoretical Physics, ETH Z\" urich, 8093 Z{\"u}rich, Switzerland}
\affiliation{Institute of Quantum Electronics, ETH Z\" urich, 8093 Z\" urich, Switzerland}
\author{Arthur Christianen}
\affiliation{Institute for Theoretical Physics, ETH Z\" urich, 8093 Z{\"u}rich, Switzerland}
\affiliation{Institute of Quantum Electronics, ETH Z\" urich, 8093 Z\" urich, Switzerland}
\author{Ataç İmamoğlu}
\affiliation{Institute of Quantum Electronics, ETH Z\" urich, 8093 Z\" urich, Switzerland}
\author{Glenn Wagner}
\affiliation{Institute for Theoretical Physics, ETH Z\" urich, 8093 Z{\"u}rich, Switzerland}

\title{Anyons at impurities: Fractional charge measurement and ``wrong sign'' binding}

\begin{abstract}
Recent photoluminescence experiments in twisted MoTe$_2$ have reported
signatures of anyons binding to trapped trions. The measured binding energies near
the fractional Chern insulators at fillings of $\nu=-2/3$ and \(-3/5\) have a ratio
close to the fractional-charge ratio \(5/3\), even though such scaling is
not generally guaranteed for the expected short-range trion-anyon interaction. We model the trapped trion as a fixed impurity, and investigate anyon-impurity binding using a
lowest-Landau-level model with exact diagonalization and Jain wavefunctions.
Surprisingly, we find that the anyon-impurity binding energy ratio between $\nu=-2/3$ and $-3/5$ is close to \(5/3\) for either
short- or long-ranged impurity-electron interactions.
However, the ratio departs strongly from the charge ratio at intermediate ranges or different Jain states.
Remarkably, we also uncover ``wrong-sign'' binding, whereby both quasiholes and quasiparticles can bind to either sign of a short-range impurity potential by selecting orbital channels with favorable local density oscillations. Our findings show that localized excitons or trions can be used to probe the internal density profile of anyons.

\end{abstract}

\maketitle

Fractionally charged anyons are a defining feature of topological order.
In fractional quantum Hall systems, fractional charge has been established
through shot-noise~\cite{DEPICCIOTTO1998395,Saminadayar1997Observation}
and antidot experiments~\cite{Goldman2001invariance}, while fractional
statistics has been probed using collider
experiments~\cite{Bartolomei2020fractional} and
interferometry~\cite{Nakamura2020anyons,Samuelson2026slow,Werkmeister2025anyon,henzinger2026controlledlocalizationanyonsgraphene}.
A major goal is to trap and manipulate individual anyons, both to
study topological phases microscopically and for potential applications in
topological quantum computing~\cite{TQC}. In conventional quantum Hall
systems, tunable trapping has been proposed in atomic
force microscopy~\cite{Hu2008trapping} and realized using scanning tunnelling
microscopy~\cite{Papic2018,park2026localspectroscopyanyonsbound,deng2026realspaceimagingquantumhall}.

Fractional Chern insulators offer a zero-field setting for studying
anyons~\cite{Neupert2011,Regnault2011,Sheng2011}. Several such states have
been observed in twisted MoTe$_2$, including states along part of the Jain
sequence at fractional fillings of \(\nu=-n/(2n-1)\)~\cite{Jain1989composite,Cai2023signatures,Zeng2023thermodynamic,Wang2025Hidden,Xu2023observation,Park2025Ferromagnetism,Ji2024Local,redekop2024directmagneticimagingfractional,Anderson2024Trion}.
Optical excitations have been proposed as local probes that can detect or
bind to anyons~\cite{mostaan2026anyon,Wagner2026sensing,huang2025thermalionizationimpurityboundquasiholes,Lu2026}.
In a recent photoluminescence experiment, spatially localized trions, formed by binding of a weakly trapped exciton to an itinerant hole, were observed~\cite{Li2026Signatures}. Upon
doping beyond the incompressible states to \(\nu=-2/3-\epsilon\) and
\(-3/5-\epsilon\),  the trapped-trion resonance jumped abruptly to lower energies; the energy difference between the trapped-trion resonances at \(\nu=-2/3\) and \(\nu=-2/3-\epsilon\) was interpreted as originating from the formation of an anyon--trion bound state.

Without access to conventional fractional-charge probes, the experiment
inferred fractionalization from the binding energies: their ratio at fillings of
\(-2/3\) and \(-3/5\) was close to the charge ratio \(5/3\).
Why the binding-energy ratio should reflect the charge ratio, however, is not obvious.
Such scaling would be natural for a broad attractive potential, which couples to an anyon’s total charge without resolving its internal structure. 
Yet the trapped trion and the anyon it binds are both positively charged, so their long-range Coulomb interaction is repulsive.
The experimental interpretation instead invokes
an attraction mediated by the neutral, tightly bound and polarizable exciton within the
trion~\cite{Li2026Signatures}.
Such an interaction would be short-ranged on the scale of the
effective magnetic length, making the binding energy sensitive to the
anyon's internal density profile rather than simply its total charge. The
observed \(5/3\) ratio thus raises the question: does impurity binding
measure fractional charge or does it probe an anyon's internal structure?

We address this question using exact diagonalization and Jain wavefunctions
for a lowest-Landau-level (LLL) model of a density-coupled impurity, varying both
the sign and range of the impurity-electron interaction potential. As expected, a long-range  potential yields binding controlled
by the fractional charge. Surprisingly, the ratio between the anyon-impurity binding energies at \(\nu=-2/3\) and \(-3/5\) also remains close to \(5/3\) in the contact limit,
although it varies strongly at intermediate ranges and departs from charge
scaling elsewhere in the Jain sequence. Strikingly, we also find
``wrong-sign'' binding: quasiholes can bind to electron-attractive impurities
and quasiparticles to electron-repulsive ones by selecting angular-momentum
channels that exploit oscillations in the internal anyon density profiles.
Impurity spectroscopy therefore measures fractional charge only in specific
regimes, while more generally probing an anyon's spatial structure.

\begin{figure}
    \centering
    \includegraphics[width=1.0\columnwidth]{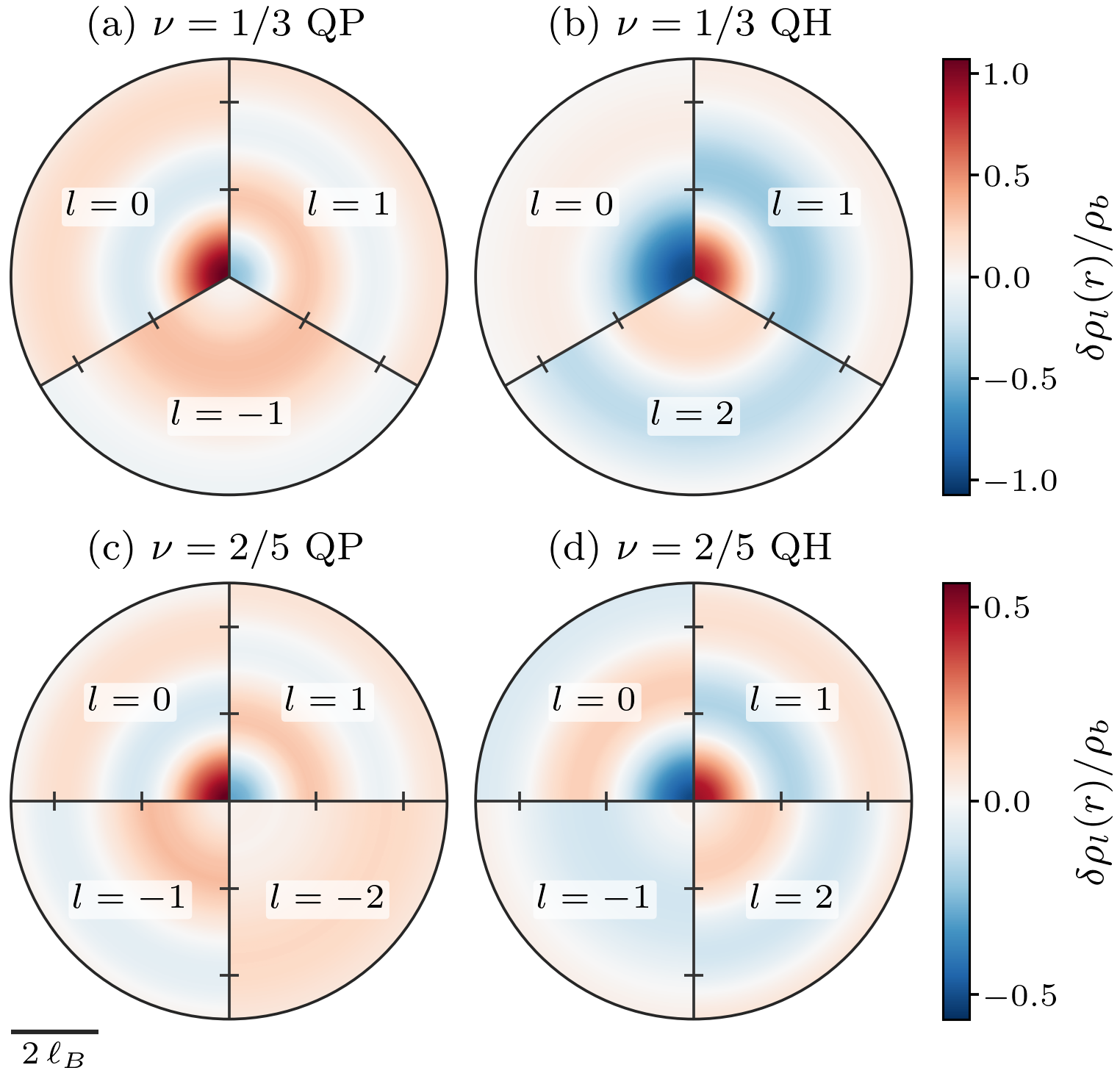}
    \caption{\textbf{Density profiles for quasiholes and quasiparticles.}
    Relative density deviation $\delta \rho_l(r) = [\rho_{s,l}(r)-\rho_b]/\rho_b$ at
    $\nu=1/3$ and $2/5$, computed by Monte Carlo sampling of Jain
    wavefunctions, with bulk density $\rho_b=\nu/(2\pi\ell_B^2)$, where $\ell_B$ is the magnetic length.
    The quasiparticle and quasihole panels contain $N_e=75$ and $73$
    electrons at $\nu=1/3$, and $N_e=91$ and $89$ at $\nu=2/5$,
    respectively. Each wedge shows the profile for an orbital label $l$
    of the composite-fermion orbital added to create a quasiparticle or
    removed to create a quasihole.}
    \label{fig:Fig2}
\end{figure}

\paragraph*{Model.---} Motivated by the recent experiment in twisted MoTe${}_2$~\cite{Li2026Signatures}, we represent the localized
optical excitation by a static impurity coupled to the carrier density.
The microscopic character of this excitation is not presently clear.
Although it is described as a localized trion in the experiment, its redshift with density may instead reflect an
exciton-polaron~\cite{sidler:2017,efimkin:2017} with a substantial excitonic component. We therefore do
not assume a microscopic trion--carrier interaction. Instead, we treat
both the sign and the spatial range of the effective impurity potential
as phenomenological parameters.

We consider spin-polarized electrons in the LLL on the
Haldane sphere, threaded by \(N_\phi=2Q\) flux quanta. The lowest Landau
level forms an angular-momentum multiplet with \(j=Q\) and
\(m=-Q,\ldots,Q\). We place the impurity at the north pole and take its
potential to be
\begin{equation}
V_a(r)=U\exp\left(-\frac{r^2}{2a^2}\right),
\label{eq:impurity-potential}
\end{equation}
where \(r\) is the chord distance from the north pole and $a$ is the Gaussian range. In our electron
convention, \(U>0\) repels electrons and \(U<0\) attracts them. By varying
\(U\) and \(a\), we can distinguish effects controlled by the net charge
of an anyon from those that depend on its internal density profile.

In the LLL orbital basis, the Hamiltonian is
\begin{equation}
H=
\sum_m U_m c_m^\dagger c_m
+\frac{1}{2}
\sum_{m_1,m_2,m_3,m_4}
V_{m_1,m_2}^{m_3,m_4}
c_{m_1}^\dagger c_{m_2}^\dagger c_{m_4}c_{m_3},
\label{eq:hamiltonian}
\end{equation}
where \(V_{m_1,m_2}^{m_3,m_4}\) are the Coulomb matrix elements and $U_m$
is the impurity-potential matrix element. Rotational symmetry about the
polar axis makes the impurity term diagonal in \(m\). This minimal model
isolates binding arising from the structure of quantum Hall
quasiparticles while neglecting Chern-band geometry, Landau-level mixing,
and the internal structure of the optical excitation.

We primarily study quasiholes and quasiparticles near electron fillings
\(\nu=1/3\) and \(2/5\). These are particle--hole conjugate to the hole
fillings studied in Ref.~\cite{Li2026Signatures}: the experimental dopings
\(\nu=-2/3-\epsilon\) and \(-3/5-\epsilon\) map respectively to
\(1/3-\epsilon\) and \(2/5-\epsilon\) in our electron convention.
Accordingly, the sign of a density-coupled potential is reversed when
translating between the hole and electron conventions.

\paragraph*{Quasihole and quasiparticle wavefunctions.---} To understand what a localized impurity probes, we first examine the density profiles of individual quasiholes and quasiparticles following the Jain construction~\cite{Jain1989composite,Jain2009CompositeFermions}.
We attach $2p$ vortices to each electron to form composite fermions. Each composite fermion sees the original
flux $N_\phi$ together with the statistical flux of the $N_e-1$ other particles, giving an effective monopole strength $2Q^* = N_\phi-2p(N_e-1)$.
The Jain wavefunction is then constructed by filling $n$ effective Landau levels ($\Lambda$ levels) at monopole strength $Q^*$ to form the Slater determinant $\Phi_n$,
attaching Jastrow factors, and projecting to the LLL:
\begin{equation}
 \Psi_{n,p}=\PLLL\left[\Phi_n J^{2p}\right].
 \label{eq:jain}
\end{equation}
Here, $\PLLL$ is LLL projection, $J=\prod_{i<j}(u_i v_j-u_j v_i)$ is the spherical Jastrow factor, and $(u_i,v_i)$ are the spinor coordinates of electron $i$.
To handle the LLL projection at large system sizes, we employ the standard approximation of Jain-Kamilla projection~\cite{JK} (see the Supplemental Material (SM) for details~\cite{supmat}).

The Jain wavefunctions capture to remarkable accuracy the FQH ground state wavefunction for the full Jain sequence $\nu = n/(2pn+1)$~\cite{Laughlin1983,Jain1989composite}. These states correspond to an integer number of filled composite fermion levels. A quasihole wavefunction can be constructed by removing one orbital in the Slater determinant from the highest filled $\Lambda$ level; a quasiparticle adds one orbital to the first empty $\Lambda$ level. The orbital that is removed or added can thus be specified by $m$. For a large sphere (i.e., the planar limit), this orbital relates to the angular momentum orbitals of the disk by $l=Q^* - m$, which characterizes the position of the anyon relative
to the north pole.

Figure~\ref{fig:Fig2} shows the electron-density profiles of quasiholes
and quasiparticles at \(\nu=1/3\) and \(2/5\) for several values of
\(l\). 
At the north pole, the \(l=0\) quasihole depletes the electron density and the \(l=0\) quasiparticle enhances it. For \(l=1\), these signs reverse: the quasihole enhances the density at the pole, while the quasiparticle depletes it.
Increasing \(l\) shifts the dominant density disturbance away
from the pole. Importantly, the density response is not monotonic:
although a quasihole carries a net electron deficit, its density profile
contains regions of enhanced electron density, and conversely for a
quasiparticle.
These oscillations imply that a local impurity does not couple only to
the integrated charge of an anyon. A short-range potential probes the
density close to the impurity, whose magnitude and sign depend on
the anyon orbital. A sufficiently broad potential instead averages over
the internal density profile and becomes sensitive to the integrated
fractional charge. These results suggest that the preferred orbital and the sign of the potential capable of binding the anyon can
depend on the impurity range.

\paragraph*{Impurity--anyon binding.---} We study binding using both exact
diagonalization (ED) of the Hamiltonian in Eq.~\eqref{eq:hamiltonian} and Jain trial
wavefunctions. Let \(E^{\rm FQH}(U)\) and \(E^s(U)\) be the energies of
the reference fluid and the sector containing one anyon of type
\(s=\mathrm{QH},\mathrm{QP}\). We define the impurity-induced binding
energy as
\begin{equation}
E_{b;s}(U)=
\bigl[E^s(U)-E^{\rm FQH}(U)\bigr]
-\bigl[E^s(0)-E^{\rm FQH}(0)\bigr],
\label{eq:binding-energy}
\end{equation}
which can be directly computed with ED.

In the Jain calculation, we solve the impurity-anyon problem in a subspace spanned by single-anyon Jain states.  For orbital \(l\), define
\(\delta\rho_{s,l}(r)=\rho_{s,l}(r)-\rho_{\rm ref}\), with
$\rho_{\rm ref}$ a uniform reference density specified below. The binding energy of this orbital is
\begin{equation}
E^{(J)}_{b;sl}(U)
=\int dA\,V_a(r)\delta\rho_{s,l}(r)
\equiv U I_{s,l}(a).
\label{eq:jain-binding-overlap}
\end{equation}
Within the single-anyon manifold, axial symmetry prevents mixing between
different \(l\), so the preferred channel minimizes \(U I_{s,l}\).
Hence \(U>0\) selects the smallest overlap and \(U<0\) the largest. The details on the Monte Carlo implementation are given in
the SM~\cite{supmat}.

\begin{figure}
    \centering
    \includegraphics[width=\columnwidth]{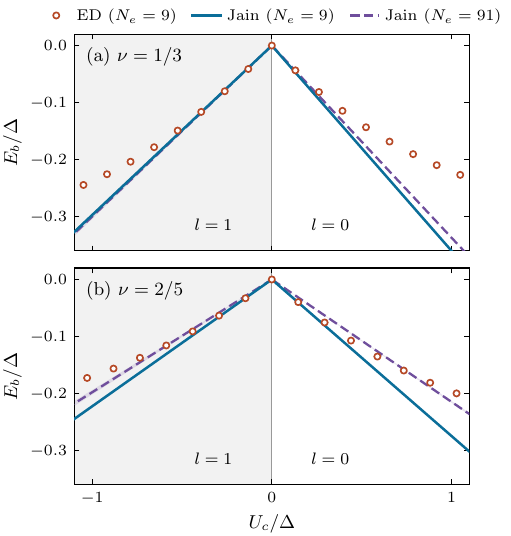}
    \caption{\textbf{Contact-impurity quasihole binding.}
    Binding energy for (a) \(\nu=1/3\) and (b) \(\nu=2/5\) quasiholes as a function of the projected contact-impurity strength \(U_c\). Open circles show ED for \(N_e=9\); solid and dashed lines show Jain Monte Carlo estimates for \(N_e=9\) and \(N_e=91\), respectively. For $N_e=9$, the Jain contact slopes subtract the north-pole orbital occupation of the corresponding incompressible ED parent; for $N_e=91$, they use the shift-corrected reference density $\rho_{\rm ref}=\nu(N_\phi+\mathcal S)/\mathcal A$ on the quasihole sphere. The dashed line for \(\nu=1/3\) and \(U_c>0\) uses the exact Laughlin quasihole contact node. In-panel labels indicate the preferred Jain orbital \(l\).}
    \label{fig:qh-binding-comparison}
\end{figure}

We first examine the contact limit, where Fig.~\ref{fig:qh-binding-comparison} reveals striking wrong-sign binding: a quasihole binds to an electron-attractive impurity almost as strongly as to an electron-repulsive one. Both ED and the Jain calculation capture this effect and agree quantitatively up to $|U_c|/\Delta_\nu\sim0.3$, where $U_c$ denotes the
coefficient of $c_Q^\dagger c_Q$ and $\Delta_\nu$ is the finite-size many-body gap.
For the $N_e=9$ Jain comparison, we choose $\rho_{\rm ref}$ so that the subtracted contact occupation matches the north-pole orbital occupation of the corresponding incompressible reference fluid in ED at $U_c=0$.
At stronger coupling, excitations across the gap allow the reference fluid to polarize --- an effect absent from the Jain estimate --- so the ED binding energies become less negative. We examine this departure and present the analogous quasiparticle comparison in the End Matter.

The Jain calculation can achieve significantly larger system sizes than ED, and we use its binding estimates in what follows in the weak-coupling regime benchmarked above. In Fig.~\ref{fig:qh-binding-comparison}, the \(N_e=91\) slopes differ from those at \(N_e=9\), most clearly for the \(\nu=2/5\) quasihole at \(U_c>0\), pointing to substantial size dependence at the system sizes accessible to ED. For these larger systems, we take \(\rho_{\rm ref}\) to be the electron density sampled by the impurity when the anyon is far away. 
For $\nu=n/(2pn+1)$, the anyon carries charge $e^*=\pm e/(2pn+1)$, where $e>0$ is the elementary charge; the plus sign denotes a quasihole and the minus sign a quasiparticle. Accounting for this localized charge, the far-field density is therefore $\rho_{\rm ref}=(N_e+e^*/e)/\mathcal A$, where $\mathcal A=2\pi N_\phi\ell_B^2$ is the area of the sphere.

\begin{figure*}
    \centering
    \includegraphics[width=\linewidth]{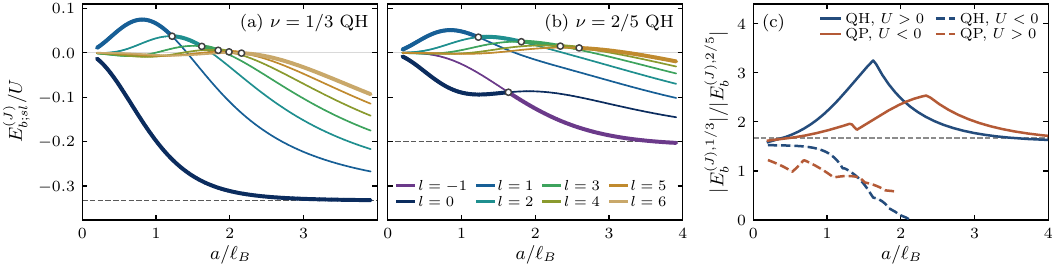}
    \caption{\textbf{Impurity overlap of Jain anyons.} (a,b) Gaussian-impurity
    binding energies $E^{(J)}_{b;sl}/U$ for quasiholes at $\nu=1/3$ and $2/5$.
    (c) Ratio of the $1/3$ and $2/5$ binding-energy magnitudes for quasiholes
    (blue) and quasiparticles (orange). Dark curves use the usual impurity sign
    ($U>0$ for quasiholes and $U<0$ for quasiparticles); dashed curves use the opposite sign.
    The wrong-sign quasiparticle ratio is
truncated when the extremal $2/5$ channel reaches the edge of the
sampled orbital ladder, beyond which its envelope is unresolved.}
    \label{fig:binding}
\end{figure*}

Figures~\ref{fig:binding}(a) and (b) show how the range of the impurity potential controls binding.
As the range varies, the orbital with the most favorable overlap with the potential changes.
At short range, we find the wrong-sign binding seen in the contact limit. As the potential extends beyond the anyon’s internal structure, this effect disappears; for $a\gtrsim 3\ell_B$, the binding energy scales with the anyon’s charge.

Figure~\ref{fig:binding}(c) shows the ratio of the \(1/3\) and \(2/5\) binding-energy magnitudes, with the lowest-energy orbital selected independently at each filling. At large potential ranges, the ratio approaches the charge ratio \(5/3\). More surprisingly, it is also close to \(5/3\) in the contact limit, consistent with the scaling observed in Ref.~\cite{Li2026Signatures}.
At intermediate ranges, however, the ratio departs appreciably from
$5/3$. Initially, the Gaussian potential underweights the outer
density depletion of the $2/5$, $l=0$ quasihole
(Fig.~\ref{fig:Fig2}), reducing its binding-energy magnitude
relative to $1/3$. Around $a\simeq1.7\ell_B$, the preferred
$2/5$ quasihole channel switches from $l=0$ to $l=-1$,
producing the peak and subsequent decline of the ratio.
Right-sign quasiparticle binding shows a similar range dependence. We also show the wrong-sign ratios; strikingly, wrong-sign quasihole binding again gives a ratio close to \(5/3\).

A simple composite-fermion picture suggests why charge
scaling might emerge in the contact limit. Before flux attachment and
LLL projection, removing (adding) a \(\Lambda\)-level
orbital at the north pole decreases (increases) the local density by
\(\rho_{\rm b}/n=|e^*|/(2\pi\ell_B^2 e)\). This argument predicts a local
density response proportional to the fractional charge. It is not,
however, an exact result for the electronic Jain wavefunction, because
both the Jastrow factors and LLL projection can modify the
density at the impurity. Nor does the picture explain wrong-sign binding, which relies on a local density response of the opposite sign.

\paragraph*{Contact response across the Jain sequence.---}
To test whether the contact-limit agreement at \(\nu=1/3\) and \(2/5\)
is generic, we evaluate the dimensionless signed contact depletion
\begin{equation}
D_{n,p}^{s} =2\pi\ell_B^2
\left[\rho_{\rm ref}-\rho_{s,l=0}(r=0)\right],
\end{equation}
where $\rho_{\rm ref}$ is the far-field density defined above.
The contact depletion fixes the short-range binding energy through
$E^{(J)}_{b;s0}\simeq-U(a^2/\ell_B^2)D_{n,p}^{s}$. The composite-fermion argument above predicts
$D_{n,p}^{\rm QH}=-D_{n,p}^{\rm QP}=|e^*|/e$.
We test the prediction in Table~\ref{tab:D} for \(n,p=1,\ldots,4\), extrapolating to the thermodynamic limit using three system sizes for each Jain state.

\begin{table}
\centering\small
\setlength{\tabcolsep}{1.5pt}
\caption{Signed contact depletion $D_{n,p}^{s}$ of the
$l=0$ quasihole (QH) and quasiparticle (QP) states in the thermodynamic limit, from weighted linear
fits in $1/N_\phi$ to three system sizes (\(N_e\approx30,60,90\)). The final block gives
$|e^*|/e=1/(2pn+1)$ for reference. Parentheses give the uncertainty propagated from
the Monte Carlo sampling errors through the fit; they do not include any
uncertainty in the fit form, which is of comparable size. The $n=1$ quasihole row is exact.}\label{tab:D}
\begin{tabular}{ccrrrr}\toprule
 & $n$ & $p=1$ & $p=2$ & $p=3$ & $p=4$ \\\midrule
QH & 1 & 0.3333 & 0.2000 & 0.1429 & 0.1111 \\
 & 2 & 0.2061(20) & 0.1435(12) & 0.1108(11) & 0.0878(11) \\
 & 3 & 0.1459(19) & 0.1115(14) & 0.0840(14) & 0.0717(15) \\
 & 4 & 0.1199(29) & 0.0874(16) & 0.0678(13) & 0.0584(25) \\\midrule
QP & 1 & $-0.3542(48)$ & $-0.2946(42)$ & $-0.2466(39)$ & $-0.2155(36)$ \\
 & 2 & $-0.2168(51)$ & $-0.1755(31)$ & $-0.1425(50)$ & $-0.1289(48)$ \\
 & 3 & $-0.1531(53)$ & $-0.1200(53)$ & $-0.1065(49)$ & $-0.0824(40)$ \\
 & 4 & $-0.1222(69)$ & $-0.0898(47)$ & $-0.0742(39)$ & $-0.0614(47)$ \\\midrule
$|e^*|/e$ & 1 & 0.3333 & 0.2000 & 0.1429 & 0.1111 \\
 & 2 & 0.2000 & 0.1111 & 0.0769 & 0.0588 \\
 & 3 & 0.1429 & 0.0769 & 0.0526 & 0.0400 \\
 & 4 & 0.1111 & 0.0588 & 0.0400 & 0.0303 \\\bottomrule
\end{tabular}
\end{table}

Only the Laughlin quasihole (\(n=1\)) satisfies the prediction exactly: its wavefunction vanishes whenever an electron sits at the pole, giving \(D_{1,p}^{\rm QH}=\nu=|e^*|/e\). For \(p=1\), both the quasihole and quasiparticle contact depletions remain close to the fractional charge. The corresponding \(1/3\)-to-\(2/5\) ratios from Table~\ref{tab:D} are \(1.62(2)\) for quasiholes and \(1.63(4)\) for quasiparticles, compared with the charge ratio \(5/3\). Both are within a few percent of this value, in line with the near-charge scaling of the experimentally relevant states in Fig.~\ref{fig:binding}(c).

Across the rest of the Jain sequence,
$|e^*|/e$ underestimates the magnitude of the contact depletion. The quasihole results nevertheless show an approximate $n\leftrightarrow p$ symmetry, as does $|e^*|/e=1/(2np+1)$. More specifically, the quasihole values appear to be organized approximately by $n+p$. We extend this analysis to wrong-sign binding in the $l=1$ channel in the End Matter. 
There, the \(p=1\) quasiholes again show approximate charge scaling, but with a contact response of the sign expected for a quasiparticle. Thus, the sign and magnitude of contact binding alone need not distinguish a quasihole from a quasiparticle.
The $l=1$ quasiparticle depletions differ more substantially in magnitude from $|e^*|/e$, yet appear approximately symmetric under $n\leftrightarrow p$; the $l=1$ quasiholes do not show this symmetry.

\paragraph*{Discussion and outlook.---}
Impurity spectroscopy measures fractional charge when the potential is
broad enough to average over the anyon's internal density profile. A
short-range potential instead probes this structure, allowing
wrong-sign binding by optimizing the orbital. At
\(\nu=1/3\) and \(2/5\), the contact-limit binding-energy ratio happens to
lie close to the charge ratio \(5/3\), showing that the experimental ratio
can also arise from a short-range interaction. This agreement is not
generic across impurity ranges or the Jain sequence. Measurements at
additional fractions
could
distinguish charge-dominated binding from sensitivity to internal
structure.

A quantitative description of the twisted MoTe$_2$ experiment requires
ingredients beyond our minimal model. At finite doping, interactions
between anyons may alter trapping and produce anyon
``molecules''~\cite{Gattu2025molecular,Xu2025dynamics,Heras2020anyonic}.
Anyon dispersion in a fractional Chern insulator also introduces a
localization cost; the larger quasiparticle dispersion predicted for
twisted MoTe$_2$ may suppress quasiparticle relative to quasihole
binding~\cite{Goncalves2026spinless}. Future work should incorporate these
effects together with the internal dynamics of the trapped optical
excitation.

\paragraph*{Note added.---}
During completion of this work, we became aware of related studies~\cite{wang2026anyonimpurityboundstatesquantumengineered,cui2026anyonicmoleculestrionlooking,LuSantos2026AnyonBoundStates}. 

\paragraph*{Acknowledgements.---} Exact diagonalization calculations were performed using DiagHam \cite{Regnault2001DiagHam}. We thank Siddharth Parameswaran for comments on a version of the manuscript. G.W.~is supported by the Swiss National Science Foundation (SNSF) via Ambizione grant number PZ00P2-216183. H.S.A., A.C. and A.I. were supported by the Swiss National Science Foundation (SNSF) under Grant No. 2000-1-240035. OpenAI ChatGPT and Anthropic Claude were used to assist with code development for figures and the implementation of Jain wave functions. All AI-assisted code and resulting calculations were independently reviewed and validated by the authors.

\bibliography{references}

@article{Saminadayar1997Observation,
  title = {Observation of the $\mathit{e}\mathit{/}3$ Fractionally Charged Laughlin Quasiparticle},
  author = {Saminadayar, L. and Glattli, D. C. and Jin, Y. and Etienne, B.},
  journal = {Phys. Rev. Lett.},
  volume = {79},
  issue = {13},
  pages = {2526--2529},
  numpages = {0},
  year = {1997},
  month = {Sep},
  publisher = {American Physical Society},
  doi = {10.1103/PhysRevLett.79.2526},
  url = {https://link.aps.org/doi/10.1103/PhysRevLett.79.2526}
}

@article{Fano1986configuration,
  title = {Configuration-interaction calculations on the fractional quantum Hall effect},
  author = {Fano, G. and Ortolani, F. and Colombo, E.},
  journal = {Phys. Rev. B},
  volume = {34},
  issue = {4},
  pages = {2670--2680},
  numpages = {0},
  year = {1986},
  month = {Aug},
  publisher = {American Physical Society},
  doi = {10.1103/PhysRevB.34.2670},
  url = {https://link.aps.org/doi/10.1103/PhysRevB.34.2670}
}

@article{Wu1977,
  title = {Some properties of monopole harmonics},
  author = {Wu, Tai Tsun and Yang, Chen Ning},
  journal = {Phys. Rev. D},
  volume = {16},
  issue = {4},
  pages = {1018--1021},
  numpages = {0},
  year = {1977},
  month = {Aug},
  publisher = {American Physical Society},
  doi = {10.1103/PhysRevD.16.1018},
  url = {https://link.aps.org/doi/10.1103/PhysRevD.16.1018}
}

@article{Neupert2011,
  title = {Fractional Quantum Hall States at Zero Magnetic Field},
  author = {Neupert, Titus and Santos, Luiz and Chamon, Claudio and Mudry, Christopher},
  journal = {Phys. Rev. Lett.},
  volume = {106},
  issue = {23},
  pages = {236804},
  numpages = {4},
  year = {2011},
  month = {Jun},
  publisher = {American Physical Society},
  doi = {10.1103/PhysRevLett.106.236804},
  url = {https://link.aps.org/doi/10.1103/PhysRevLett.106.236804}
}

@article{Sheng2011,
	author = {Sheng, D. N. and Gu, Zheng-Cheng and Sun, Kai and Sheng, L.},
	date = {2011/07/12},
	doi = {10.1038/ncomms1380},
	id = {Sheng2011},
	isbn = {2041-1723},
	journal = {Nature Communications},
	number = {1},
	pages = {389},
	title = {Fractional quantum Hall effect in the absence of Landau levels},
	url = {https://doi.org/10.1038/ncomms1380},
	volume = {2},
	year = {2011}}

@article{Regnault2011,
  title = {Fractional Chern Insulator},
  author = {Regnault, N. and Bernevig, B. Andrei},
  journal = {Phys. Rev. X},
  volume = {1},
  issue = {2},
  pages = {021014},
  numpages = {14},
  year = {2011},
  month = {Dec},
  publisher = {American Physical Society},
  doi = {10.1103/PhysRevX.1.021014},
  url = {https://link.aps.org/doi/10.1103/PhysRevX.1.021014}
}

@article{TQC,
  title = {Non-Abelian anyons and topological quantum computation},
  author = {Nayak, Chetan and Simon, Steven H. and Stern, Ady and Freedman, Michael and Das Sarma, Sankar},
  journal = {Rev. Mod. Phys.},
  volume = {80},
  issue = {3},
  pages = {1083--1159},
  numpages = {0},
  year = {2008},
  month = {Sep},
  publisher = {American Physical Society},
  doi = {10.1103/RevModPhys.80.1083},
  url = {https://link.aps.org/doi/10.1103/RevModPhys.80.1083}
}

@misc{wang2026anyonimpurityboundstatesquantumengineered,
      title={Anyon-Impurity Bound States in Quantum-Engineered Fractional Chern Insulators}, 
      author={Botao Wang and Amit Vashisht and Felix A. Palm and Fabian Grusdt and Laurens Vanderstraeten and Nathan Goldman},
      year={2026},
      eprint={2608.06233},
      archivePrefix={arXiv},
      primaryClass={cond-mat.mes-hall},
      url={https://arxiv.org/abs/2608.06233}, 
}

@article{JK,
	author = {Jain, J. K. and Kamilla, R. K.},
	doi = {10.1142/S0217979297001301},
	eprint = {https://doi.org/10.1142/S0217979297001301},
	journal = {International Journal of Modern Physics B},
	number = {22},
	pages = {2621-2660},
	title = {Composite Fermions in the Hilbert Space of the Lowest Electronic Landau Level},
	url = {https://doi.org/10.1142/S0217979297001301},
	volume = {11},
	year = {1997}}

@article{Xu2025dynamics,
  title = {Dynamics of anyon clusters in fractional quantum Hall fluids},
  author = {Xu, Qianhui and Ji, Guangyue and Wang, Yuzhu and Trung, Ha Quang and Yang, Bo},
  journal = {Phys. Rev. B},
  volume = {112},
  issue = {23},
  pages = {235112},
  numpages = {11},
  year = {2025},
  month = {Dec},
  publisher = {American Physical Society},
  doi = {10.1103/vgz6-z98r},
  url = {https://link.aps.org/doi/10.1103/vgz6-z98r}
}

@article{Heras2020anyonic,
  title = {Anyonic Molecules in Atomic Fractional Quantum Hall Liquids: A Quantitative Probe of Fractional Charge and Anyonic Statistics},
  author = {Mu\~noz de las Heras, A. and Macaluso, E. and Carusotto, I.},
  journal = {Phys. Rev. X},
  volume = {10},
  issue = {4},
  pages = {041058},
  numpages = {23},
  year = {2020},
  month = {Dec},
  publisher = {American Physical Society},
  doi = {10.1103/PhysRevX.10.041058},
  url = {https://link.aps.org/doi/10.1103/PhysRevX.10.041058}
}

@article{Gattu2025molecular,
  title = {Molecular Anyons in the Fractional Quantum Hall Effect},
  author = {Gattu, Mytraya and Jain, J. K.},
  journal = {Phys. Rev. Lett.},
  volume = {135},
  issue = {23},
  pages = {236601},
  numpages = {7},
  year = {2025},
  month = {Dec},
  publisher = {American Physical Society},
  doi = {10.1103/scl5-8pv6},
  url = {https://link.aps.org/doi/10.1103/scl5-8pv6}
}

@article{Hu2008trapping,
  title = {Trapping Abelian anyons in fractional quantum Hall droplets},
  author = {Hu, Zi-Xiang and Wan, Xin and Schmitteckert, Peter},
  journal = {Phys. Rev. B},
  volume = {77},
  issue = {7},
  pages = {075331},
  numpages = {7},
  year = {2008},
  month = {Feb},
  publisher = {American Physical Society},
  doi = {10.1103/PhysRevB.77.075331},
  url = {https://link.aps.org/doi/10.1103/PhysRevB.77.075331}
}

@misc{deng2026realspaceimagingquantumhall,
      title={Real-space Imaging of Quantum Hall Quasiparticles}, 
      author={Jinghao Deng and Yiming Sun and Dimitri Pimenov and Takashi Taniguchi and Kenji Watanabe and Erich J Mueller and Xiaomeng Liu},
      year={2026},
      eprint={2606.25036},
      archivePrefix={arXiv},
      primaryClass={cond-mat.mes-hall},
      url={https://arxiv.org/abs/2606.25036}, 
}

@misc{cui2026anyonicmoleculestrionlooking,
      title={Anyonic molecules through the trion looking glass}, 
      author={Hao-Ran Cui and Vladimir Calvera and Umang Mehta and Hart Goldman},
      year={2026},
      eprint={2609.30380},
      archivePrefix={arXiv},
      primaryClass={cond-mat.mes-hall},
      url={https://arxiv.org/abs/2609.30380}, 
}

@misc{Regnault2001DiagHam,

  author       = {N. Regnault},

  title        = {{DiagHam}},

  year         = {2001},

  howpublished = {\url{http://www.nick-ux.org/diagham}},

  note         = {Code repository}

}

@misc{huang2025thermalionizationimpurityboundquasiholes,
      title={Thermal ionization of impurity-bound quasiholes in the fractional quantum Hall effect}, 
      author={Ke Huang and Sankar Das Sarma and Xiao Li},
      year={2025},
      eprint={2512.07769},
      archivePrefix={arXiv},
      primaryClass={cond-mat.mes-hall},
      url={https://arxiv.org/abs/2512.07769}, 
}

@article{WU1976,
	author = {Tai Tsun Wu and Chen Ning Yang},
	doi = {https://doi.org/10.1016/0550-3213(76)90143-7},
	issn = {0550-3213},
	journal = {Nuclear Physics B},
	number = {3},
	pages = {365-380},
	title = {Dirac monopole without strings: Monopole harmonics},
	url = {https://www.sciencedirect.com/science/article/pii/0550321376901437},
	volume = {107},
	year = {1976}}

@article{Wen1992shift,
  title = {Shift and spin vector: New topological quantum numbers for the Hall fluids},
  author = {Wen, X. G. and Zee, A.},
  journal = {Phys. Rev. Lett.},
  volume = {69},
  issue = {6},
  pages = {953--956},
  numpages = {0},
  year = {1992},
  month = {Aug},
  publisher = {American Physical Society},
  doi = {10.1103/PhysRevLett.69.953},
  url = {https://link.aps.org/doi/10.1103/PhysRevLett.69.953}
}

@article{DEPICCIOTTO1998395,
	author = {R de-Picciotto and M Reznikov and M Heiblum and V Umansky and G Bunin and D Mahalu},
	doi = {https://doi.org/10.1016/S0921-4526(98)00139-2},
	issn = {0921-4526},
	journal = {Physica B: Condensed Matter},
	pages = {395-400},
	title = {Direct observation of a fractional charge},
	url = {https://www.sciencedirect.com/science/article/pii/S0921452698001392},
	volume = {249-251},
	year = {1998}}

@article{Nakamura2020anyons,
	author = {Nakamura, J. and Liang, S. and Gardner, G. C. and Manfra, M. J.},
	date = {2020/09/01},
	doi = {10.1038/s41567-020-1019-1},
	id = {Nakamura2020},
	isbn = {1745-2481},
	journal = {Nature Physics},
	number = {9},
	pages = {931--936},
	title = {Direct observation of anyonic braiding statistics},
	url = {https://doi.org/10.1038/s41567-020-1019-1},
	volume = {16},
	year = {2020}}

@article{Bartolomei2020fractional,
	author = {H. Bartolomei and M. Kumar and R. Bisognin and A. Marguerite and J.-M. Berroir and E. Bocquillon and B. Pla{\c c}ais and A. Cavanna and Q. Dong and U. Gennser and Y. Jin and G. F{\`e}ve},
	doi = {10.1126/science.aaz5601},
	journal = {Science},
	number = {6487},
	pages = {173-177},
	title = {Fractional statistics in anyon collisions},
	url = {https://www.science.org/doi/abs/10.1126/science.aaz5601},
	volume = {368},
	year = {2020}}

@article{Li2026Signatures,
	author = {Li, Weijie and Wang Beach, Christiano and Hu, Chaowei and Taniguchi, Takashi and Watanabe, Kenji and Chu, Jiun-Haw and Imamo{\u g}lu, Ata{\c c} and Cao, Ting and Xiao, Di and Xu, Xiaodong},
	date = {2026/03/01},
	doi = {10.1038/s41586-026-10101-w},
	id = {Li2026},
	isbn = {1476-4687},
	journal = {Nature},
	number = {8104},
	pages = {48--53},
	title = {Signatures of fractional charges via anyon--trions in twisted MoTe2},
	url = {https://doi.org/10.1038/s41586-026-10101-w},
	volume = {651},
	year = {2026}}

@article{Papic2018,
  title = {Imaging Anyons with Scanning Tunneling Microscopy},
  author = {Papi\ifmmode \acute{c}\else \'{c}\fi{}, Zlatko and Mong, Roger S. K. and Yazdani, Ali and Zaletel, Michael P.},
  journal = {Phys. Rev. X},
  volume = {8},
  issue = {1},
  pages = {011037},
  numpages = {8},
  year = {2018},
  month = {Mar},
  publisher = {American Physical Society},
  doi = {10.1103/PhysRevX.8.011037},
  url = {https://link.aps.org/doi/10.1103/PhysRevX.8.011037}
}

@misc{park2026localspectroscopyanyonsbound,
      title={Local spectroscopy of anyons bound to charge traps}, 
      author={Jeong Min Park and Cristian Voinea and Yen-Chen Tsui and Songyang Pu and Kenji Watanabe and Takashi Taniguchi and Nigel R. Cooper and Michael P. Zaletel and Zlatko Papić and Ali Yazdani},
      year={2026},
      eprint={2606.25024},
      archivePrefix={arXiv},
      primaryClass={cond-mat.mes-hall},
      url={https://arxiv.org/abs/2606.25024}, 
}

@article{Cai2023signatures,
	author = {Cai, Jiaqi and Anderson, Eric and Wang, Chong and Zhang, Xiaowei and Liu, Xiaoyu and Holtzmann, William and Zhang, Yinong and Fan, Fengren and Taniguchi, Takashi and Watanabe, Kenji and Ran, Ying and Cao, Ting and Fu, Liang and Xiao, Di and Yao, Wang and Xu, Xiaodong},
	date = {2023/10/01},
	doi = {10.1038/s41586-023-06289-w},
	id = {Cai2023},
	isbn = {1476-4687},
	journal = {Nature},
	number = {7981},
	pages = {63--68},
	title = {Signatures of fractional quantum anomalous Hall states in twisted MoTe2},
	url = {https://doi.org/10.1038/s41586-023-06289-w},
	volume = {622},
	year = {2023}}

@article{Wang2025Hidden,
	author = {Wang, Yiping and Choe, Jeongheon and Anderson, Eric and Li, Weijie and Ingham, Julian and Arsenault, Eric A. and Li, Yiliu and Hu, Xiaodong and Taniguchi, Takashi and Watanabe, Kenji and Roy, Xavier and Basov, Dmitri and Xiao, Di and Queiroz, Raquel and Hone, James C. and Xu, Xiaodong and Zhu, X. -Y.},
	date = {2025/05/01},
	doi = {10.1038/s41586-025-08954-8},
	id = {Wang2025},
	isbn = {1476-4687},
	journal = {Nature},
	number = {8065},
	pages = {1149--1155},
	title = {Hidden states and dynamics of fractional fillings in twisted MoTe2 bilayers},
	url = {https://doi.org/10.1038/s41586-025-08954-8},
	volume = {641},
	year = {2025}}

@article{Zeng2023thermodynamic,
	author = {Zeng, Yihang and Xia, Zhengchao and Kang, Kaifei and Zhu, Jiacheng and Kn{\"u}ppel, Patrick and Vaswani, Chirag and Watanabe, Kenji and Taniguchi, Takashi and Mak, Kin Fai and Shan, Jie},
	date = {2023/10/01},
	doi = {10.1038/s41586-023-06452-3},
	id = {Zeng2023},
	isbn = {1476-4687},
	journal = {Nature},
	number = {7981},
	pages = {69--73},
	title = {Thermodynamic evidence of fractional Chern insulator in moir{\'e}MoTe2},
	url = {https://doi.org/10.1038/s41586-023-06452-3},
	volume = {622},
	year = {2023}}

@article{Xu2023observation,
  title = {Observation of Integer and Fractional Quantum Anomalous Hall Effects in Twisted Bilayer ${\mathrm{MoTe}}_{2}$},
  author = {Xu, Fan and Sun, Zheng and Jia, Tongtong and Liu, Chang and Xu, Cheng and Li, Chushan and Gu, Yu and Watanabe, Kenji and Taniguchi, Takashi and Tong, Bingbing and Jia, Jinfeng and Shi, Zhiwen and Jiang, Shengwei and Zhang, Yang and Liu, Xiaoxue and Li, Tingxin},
  journal = {Phys. Rev. X},
  volume = {13},
  issue = {3},
  pages = {031037},
  numpages = {12},
  year = {2023},
  month = {Sep},
  publisher = {American Physical Society},
  doi = {10.1103/PhysRevX.13.031037},
  url = {https://link.aps.org/doi/10.1103/PhysRevX.13.031037}
}

@article{Park2025Ferromagnetism,
	author = {Park, Heonjoon and Cai, Jiaqi and Anderson, Eric and Zhang, Xiao-Wei and Liu, Xiaoyu and Holtzmann, William and Li, Weijie and Wang, Chong and Hu, Chaowei and Zhao, Yuzhou and Taniguchi, Takashi and Watanabe, Kenji and Yang, Jihui and Cobden, David and Chu, Jiun-haw and Regnault, Nicolas and Bernevig, B. Andrei and Fu, Liang and Cao, Ting and Xiao, Di and Xu, Xiaodong},
	date = {2025/04/01},
	doi = {10.1038/s41567-025-02804-0},
	id = {Park2025},
	isbn = {1745-2481},
	journal = {Nature Physics},
	number = {4},
	pages = {549--555},
	title = {Ferromagnetism and topology of the higher flat band in a fractional Chern insulator},
	url = {https://doi.org/10.1038/s41567-025-02804-0},
	volume = {21},
	year = {2025}}

@article{Ji2024Local,
	author = {Ji, Zhurun and Park, Heonjoon and Barber, Mark E. and Hu, Chaowei and Watanabe, Kenji and Taniguchi, Takashi and Chu, Jiun-Haw and Xu, Xiaodong and Shen, Zhi-Xun},
	date = {2024/11/01},
	doi = {10.1038/s41586-024-08092-7},
	id = {Ji2024},
	isbn = {1476-4687},
	journal = {Nature},
	number = {8039},
	pages = {578--583},
	title = {Local probe of bulk and edge states in a fractional Chern insulator},
	url = {https://doi.org/10.1038/s41586-024-08092-7},
	volume = {635},
	year = {2024}}

@misc{redekop2024directmagneticimagingfractional,
      title={Direct magnetic imaging of fractional Chern insulators in twisted MoTe$_2$ with a superconducting sensor}, 
      author={Evgeny Redekop and Canxun Zhang and Heonjoon Park and Jiaqi Cai and Eric Anderson and Owen Sheekey and Trevor Arp and Grigory Babikyan and Samuel Salters and Kenji Watanabe and Takashi Taniguchi and Xiaodong Xu and Andrea F. Young},
      year={2024},
      eprint={2405.10269},
      archivePrefix={arXiv},
      primaryClass={cond-mat.mes-hall},
      url={https://arxiv.org/abs/2405.10269}, 
}

@article{Anderson2024Trion,
	author = {Anderson, Eric and Cai, Jiaqi and Reddy, Aidan P. and Park, Heonjoon and Holtzmann, William and Davis, Kai and Taniguchi, Takashi and Watanabe, Kenji and Smolenski, Tomasz and Imamo{\u g}lu, Ata{\c c} and Cao, Ting and Xiao, Di and Fu, Liang and Yao, Wang and Xu, Xiaodong},
	date = {2024/11/01},
	doi = {10.1038/s41586-024-08134-0},
	id = {Anderson2024},
	isbn = {1476-4687},
	journal = {Nature},
	number = {8039},
	pages = {590--595},
	title = {Trion sensing of a zero-field composite Fermi liquid},
	url = {https://doi.org/10.1038/s41586-024-08134-0},
	volume = {635},
	year = {2024}}

@article{mostaan2026anyon,
  title = {Anyon-Trions in Atomically Thin Semiconductor Heterostructures},
  author = {Mostaan, Nader and Goldman, Nathan and Imamo{\u g}lu, Ata{\c c} and Grusdt, Fabian},
  journal = {PRX Quantum},
  volume = {7},
  issue = {1},
  pages = {010325},
  numpages = {17},
  year = {2026},
  month = {Feb},
  publisher = {American Physical Society},
  doi = {10.1103/hxmb-pn4z},
  url = {https://link.aps.org/doi/10.1103/hxmb-pn4z}
}

@article{Wagner2026sensing,
  title = {Sensing the binding and unbinding of anyons at impurities},
  author = {Wagner, Glenn and Neupert, Titus},
  journal = {Phys. Rev. Res.},
  volume = {8},
  issue = {1},
  pages = {013263},
  numpages = {9},
  year = {2026},
  month = {Mar},
  publisher = {American Physical Society},
  doi = {10.1103/cfm5-wgz1},
  url = {https://link.aps.org/doi/10.1103/cfm5-wgz1}
}

@article{Haldane1983,
  title = {Fractional Quantization of the Hall Effect: A Hierarchy of Incompressible Quantum Fluid States},
  author = {Haldane, F. D. M.},
  journal = {Phys. Rev. Lett.},
  volume = {51},
  issue = {7},
  pages = {605--608},
  numpages = {0},
  year = {1983},
  month = {Aug},
  publisher = {American Physical Society},
  doi = {10.1103/PhysRevLett.51.605},
  url = {https://link.aps.org/doi/10.1103/PhysRevLett.51.605}
}

@article{Laughlin1983,
  title = {Anomalous Quantum Hall Effect: An Incompressible Quantum Fluid with Fractionally Charged Excitations},
  author = {Laughlin, R. B.},
  journal = {Phys. Rev. Lett.},
  volume = {50},
  issue = {18},
  pages = {1395--1398},
  numpages = {0},
  year = {1983},
  month = {May},
  publisher = {American Physical Society},
  doi = {10.1103/PhysRevLett.50.1395},
  url = {https://link.aps.org/doi/10.1103/PhysRevLett.50.1395}
}

@article{Jain1989composite,
  title = {Composite-fermion approach for the fractional quantum Hall effect},
  author = {Jain, J. K.},
  journal = {Phys. Rev. Lett.},
  volume = {63},
  issue = {2},
  pages = {199--202},
  numpages = {0},
  year = {1989},
  month = {Jul},
  publisher = {American Physical Society},
  doi = {10.1103/PhysRevLett.63.199},
  url = {https://link.aps.org/doi/10.1103/PhysRevLett.63.199}
}

@article{Goncalves2026spinless,
  title = {Spinless and Spinful Charge Excitations in Moir\'e Fractional Chern Insulators},
  author = {Gon\ifmmode \mbox{\c{c}}\else \c{c}\fi{}alves, Miguel and Mendez-Valderrama, Juan Felipe and Herzog-Arbeitman, Jonah and Yu, Jiabin and Xu, Xiaodong and Xiao, Di and Bernevig, B. Andrei and Regnault, Nicolas},
  journal = {Phys. Rev. Lett.},
  volume = {136},
  issue = {19},
  pages = {196503},
  numpages = {8},
  year = {2026},
  month = {May},
  publisher = {American Physical Society},
  doi = {10.1103/gmkp-bxf2},
  url = {https://link.aps.org/doi/10.1103/gmkp-bxf2}
}

@misc{henzinger2026controlledlocalizationanyonsgraphene,
      title={Controlled localization of anyons in a graphene quantum Hall interferometer}, 
      author={Christina E. Henzinger and James R. Ehrets and Rikuto Fushio and Junkai Dong and Thomas Werkmeister and Marie E. Wesson and Kenji Watanabe and Takashi Taniguchi and Ashvin Vishwanath and Bertrand I. Halperin and Amir Yacoby and Philip Kim},
      year={2026},
      eprint={2603.11182},
      archivePrefix={arXiv},
      primaryClass={cond-mat.mes-hall},
      url={https://arxiv.org/abs/2603.11182}, 
}

@article{Goldman2001invariance,
  title = {Invariance of charge of Laughlin quasiparticles},
  author = {Goldman, V. J. and Karakurt, I. and Liu, Jun and Zaslavsky, A.},
  journal = {Phys. Rev. B},
  volume = {64},
  issue = {8},
  pages = {085319},
  numpages = {5},
  year = {2001},
  month = {Aug},
  publisher = {American Physical Society},
  doi = {10.1103/PhysRevB.64.085319},
  url = {https://link.aps.org/doi/10.1103/PhysRevB.64.085319}
}

@article{Werkmeister2025anyon,
	author = {Thomas Werkmeister and James R. Ehrets and Marie E. Wesson and Danial H. Najafabadi and Kenji Watanabe and Takashi Taniguchi and Bertrand I. Halperin and Amir Yacoby and Philip Kim},
	doi = {10.1126/science.adp5015},
	journal = {Science},
	number = {6748},
	pages = {730-735},
	title = {Anyon braiding and telegraph noise in a graphene interferometer},
	url = {https://www.science.org/doi/abs/10.1126/science.adp5015},
	volume = {388},
	year = {2025}}

@article{Samuelson2026slow,
  title = {Slow Quasiparticle Dynamics and Anyonic Statistics in a Fractional Quantum Hall Fabry-P\'erot Interferometer},
  author = {Samuelson, Noah L. and Cohen, Liam A. and Wang, Will and Blanch, Simon and Taniguchi, Takashi and Watanabe, Kenji and Zaletel, Michael P. and Young, Andrea F.},
  journal = {Phys. Rev. X},
  volume = {16},
  issue = {1},
  pages = {011062},
  numpages = {12},
  year = {2026},
  month = {Mar},
  publisher = {American Physical Society},
  doi = {10.1103/fwjg-mx9h},
  url = {https://link.aps.org/doi/10.1103/fwjg-mx9h}
}

@book{Jain2009CompositeFermions,
  author    = {Jain, Jainendra K.},
  title     = {Composite Fermions},
  publisher = {Cambridge University Press},
  year      = {2009},
  isbn      = {9780511607561},
  doi       = {10.1017/CBO9780511607561}
}

@misc{supmat,
  note = {See the Supplemental Material for details of the spherical geometry, exact diagonalization, Jain wave functions and projection, Monte Carlo methods, and finite-size extrapolation of the contact depletion. This includes Refs.~\cite{Haldane1983,WU1976,Wu1977,Fano1986configuration,Wen1992shift}.}
}

@article{sidler:2017,
  title={Fermi polaron-polaritons in charge-tunable atomically thin semiconductors},
  author={Sidler, Meinrad and Back, Patrick and Cotlet, Ovidiu and Srivastava, Ajit and Fink, Thomas and Kroner, Martin and Demler, Eugene and Imamoglu, Atac},
  journal={Nat. Phys.},
  volume={13},
  number={3},
  pages={255--261},
  year={2017},
  publisher={Nature Publishing Group UK London},
doi={10.1038/nphys3949}
}

@article{efimkin:2017,
  title = {Many-body theory of trion absorption features in two-dimensional semiconductors},
  author = {Efimkin, Dmitry K. and MacDonald, Allan H.},
  journal = {Phys. Rev. B},
  volume = {95},
  issue = {3},
  pages = {035417},
  numpages = {10},
  year = {2017},
  month = {Jan},
  publisher = {American Physical Society},
  doi = {10.1103/PhysRevB.95.035417},
  url = {https://link.aps.org/doi/10.1103/PhysRevB.95.035417}
}

@article{Lu2026,
  title = {Exciton-Anyon Binding in Fractional Chern Insulators: Spectral Fingerprints},
  author = {Lu, Tianhong and Santos, Luiz H.},
  journal = {Phys. Rev. Lett.},
  volume = {136},
  issue = {25},
  pages = {256605},
  numpages = {7},
  year = {2026},
  month = {Jun},
  publisher = {American Physical Society},
  doi = {10.1103/8bp8-g8g4},
  url = {https://link.aps.org/doi/10.1103/8bp8-g8g4}
}

@misc{LuSantos2026AnyonBoundStates,
  author = {T. Lu and L. H. Santos},
  title = {Anyon bound states in {MoTe}$_2$ fractional Chern insulator},
  year = {2026},
  note = {arXiv:2610.xxxxx}
}

\clearpage
\onecolumngrid

\section*{END MATTER}

\twocolumngrid

\begin{figure}
    \centering
    \includegraphics[width=\columnwidth]{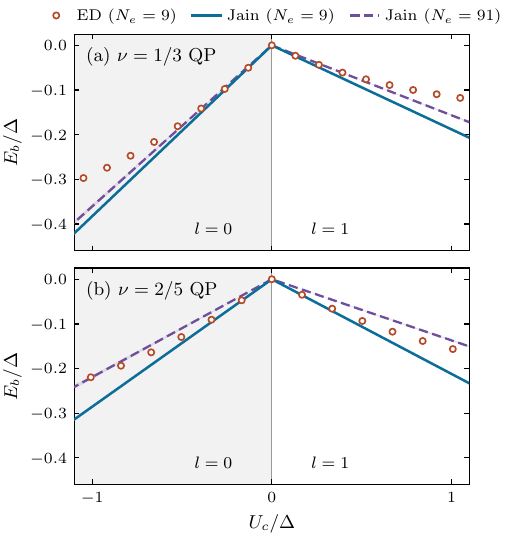}
    \caption{\textbf{Quasiparticle binding to a contact impurity.}
    Binding energy at (a) $\nu=1/3$ and (b) $\nu=2/5$, with each axis
    scaled by that filling's many-body gap $\Delta_\nu$. Open circles are
    ED results for $N_e=9$; solid lines are matched-size Jain estimates; dashed lines use $N_e=91$ Jain states.}
    \label{fig:qp-binding}
\end{figure}

\begin{figure}
    \centering
    \includegraphics[width=\columnwidth]{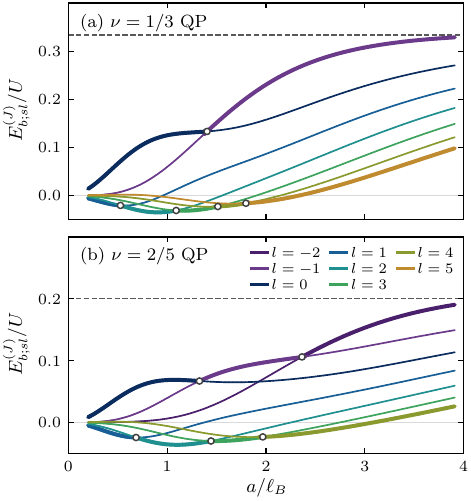}
    \caption{\textbf{Quasiparticle binding across impurity ranges.}
    Gaussian impurity binding energies $E^{(J)}_{b;sl}/U$ for quasiparticles at (a) $\nu=1/3$
    and (b) $\nu=2/5$, using the same reference density as in
    Fig.~\ref{fig:binding}. The curves show the energy of individual orbitals;
    heavy segments follow the extremal channel, and open circles mark
    changes in the preferred orbital channel.}
    \label{fig:qp-overlap}
\end{figure}

\paragraph*{Quasiparticle analysis.---} Here, we present the quasiparticle counterparts to the quasihole figures shown in the main text. The comparison of exact diagonalization and the Jain estimate is shown in Fig.~\ref{fig:qp-binding}. For $U_c<0$ the preferred Jain orbital is
$l=0$, whereas for $U_c>0$ it is $l=1$. The latter is the
wrong-sign branch. 
Figure~\ref{fig:qp-overlap} resolves the quasiparticle Jain binding
coefficient by orbital and impurity range. We observe that the wrong-sign binding is only possible for a short-range impurity potential.

\paragraph*{Density profiles with exact diagonalization.---} To elucidate the difference between the Jain estimate and the exact diagonalization, we show the density profiles of the fractional quantum Hall state, as well as the quasihole and quasiparticle states, as calculated by exact diagonalization in the presence of the short-range impurity. We present the results for the Laughlin $1/3$ state.

\begin{figure}
    \centering
    \includegraphics[width=\linewidth]{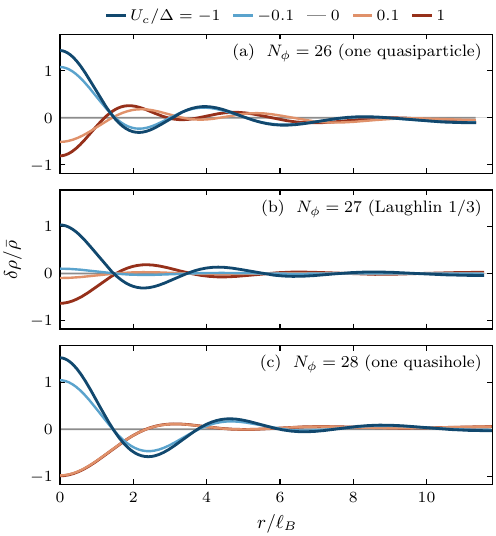}
    \caption{\textbf{Density profiles for short-range impurity.} We show the exact diagonalization density profiles with $N_e=10$ particles for the quasiparticle, FQH and quasihole sectors, respectively. We show the results for five values of $U_c/\Delta$, where $U_c$ is the strength of the contact impurity potential and $\Delta$ is the neutral gap of the FQH state.}
    \label{fig:density_profiles}
\end{figure}

As can be seen from Fig.~\ref{fig:density_profiles}, in the absence of an impurity, the FQH sector has a uniform density. There is a unique ground state with gap $\Delta$. In the quasihole and quasiparticle sectors, there are a number of exactly degenerate ground states in the absence of the impurity; averaging over these degenerate ground states, one also obtains a uniform density. Adding a weak impurity $|U_c|/\Delta=0.1$ allows the quasihole and quasiparticle to adjust by selecting the state within the degenerate ground state manifold with the most ideal location of the anyon for the given impurity. On the other hand, the response of the FQH state only appears in perturbation theory and is hence suppressed by the gap, $\delta\rho/\bar\rho\sim(|U_c|/\Delta)$. For $|U_c|/\Delta=1$ the FQH state strongly responds to the impurity. This causes the deviation from the linear behaviour in Fig.~\ref{fig:qh-binding-comparison}. Rewriting Eq.~\eqref{eq:binding-energy} as
\begin{equation}
E_{b;s}(U)=
\bigl[E^s(U)-E^s(0)\bigr]
-\bigl[E^{\rm FQH}(U)-E^{\rm FQH}(0)\bigr],
\end{equation}
the polarization of the FQH state at large $|U_c|/\Delta$ causes $\bigl[E^{\rm FQH}(U)-E^{\rm FQH}(0)\bigr]$ to become more negative, hence making $E_{b;s}(U)$ in ED less negative than in the Jain estimate (which neglects the polarization of the FQH state).

For a Coulomb interaction Hamiltonian, the Jain $2/5$ state has a smaller gap $\Delta$ than the Laughlin $1/3$ state. Hence for a given impurity strength, the Jain $2/5$ state responds more strongly than the Laughlin $1/3$ state.

\paragraph*{Wrong-sign binding contact depletion.---}

To quantify wrong-sign contact binding across the Jain sequence, we repeat the Table~\ref{tab:D} analysis for orbital $l=1$. We define $D_{n,p}^{s,l=1}=2\pi\ell_B^2[\rho_{\rm ref}-\rho_{s,l=1}(0)]$, so, as in the main text, $E^{(J)}_{b;s1}\simeq-U(a^2/\ell_B^2)D_{n,p}^{s,l=1}$ in the short-range limit. Table~\ref{tab:D-wrongsign} gives the thermodynamic intercepts of weighted fits linear in $1/N_\phi$ at three system sizes. The quasihole depletions are negative and remain close in magnitude to $|e^*|/e$ along $p=1$, whereas the positive quasiparticle depletions show much stronger deviations across the whole sequence. These signs imply binding for $U_c<0$ and $U_c>0$, respectively. Unlike the $l=0$ quasiholes, the $l=1$ quasiholes do not show approximate $n\leftrightarrow p$ symmetry. The $l=1$ quasiparticles, however, appear roughly symmetric under this exchange.

\begin{table}
\centering\small
\setlength{\tabcolsep}{1.5pt}
\caption{Signed contact depletion $D_{n,p}^{s,l=1}$ of the wrong-sign
$l=1$ quasihole (QH) and quasiparticle (QP) states in the thermodynamic
limit. The weighted linear fit and uncertainty analysis is the same as in Table~\ref{tab:D}.}
\label{tab:D-wrongsign}
\begin{tabular}{ccrrrr}\toprule
 & $n$ & $p=1$ & $p=2$ & $p=3$ & $p=4$ \\\midrule
QH & 1 & $-0.2954(48)$ & $-0.2615(41)$ & $-0.2266(37)$ & $-0.1916(28)$ \\
 & 2 & $-0.1889(35)$ & $-0.1498(31)$ & $-0.1206(29)$ & $-0.0996(43)$ \\
 & 3 & $-0.1393(30)$ & $-0.1025(31)$ & $-0.0857(28)$ & $-0.0698(35)$ \\
 & 4 & $-0.1064(35)$ & $-0.0788(33)$ & $-0.0535(33)$ & $-0.0556(38)$ \\\midrule
QP & 1 & $0.1534(22)$ & $0.1228(13)$ & $0.0995(10)$ & $0.0839(8)$ \\
 & 2 & $0.1237(34)$ & $0.0931(23)$ & $0.0730(26)$ & $0.0612(15)$ \\
 & 3 & $0.1054(39)$ & $0.0733(31)$ & $0.0585(24)$ & $0.0460(24)$ \\
 & 4 & $0.0870(33)$ & $0.0568(26)$ & $0.0473(37)$ & $0.0410(30)$ \\\midrule
$|e^*|/e$ & 1 & 0.3333 & 0.2000 & 0.1429 & 0.1111 \\
 & 2 & 0.2000 & 0.1111 & 0.0769 & 0.0588 \\
 & 3 & 0.1429 & 0.0769 & 0.0526 & 0.0400 \\
 & 4 & 0.1111 & 0.0588 & 0.0400 & 0.0303 \\\bottomrule
\end{tabular}
\end{table}
\clearpage
\renewcommand{\theequation}{S\arabic{equation}}
\renewcommand{\thefigure}{S\arabic{figure}}
\renewcommand{\thetable}{S\arabic{table}}
\renewcommand{\theHequation}{S\arabic{equation}}
\renewcommand{\theHfigure}{S\arabic{figure}}
\renewcommand{\theHtable}{S\arabic{table}}


\setcounter{equation}{0}
\setcounter{figure}{0}
\setcounter{table}{0}

\clearpage
\onecolumngrid
\section*{SUPPLEMENTAL MATERIAL:\\ ``Anyons at impurities: Fractional charge measurement and ``wrong sign'' binding''}
\setcounter{page}{1}
\begin{center}
Haydn S. Adlong$^{1,2}$, Arthur Christianen$^{1,2}$,
Ata\c{c} \.{I}mamo\u{g}lu$^{2}$, and Glenn Wagner$^{1}$\\[0.5em]
{\small $^1$Institute for Theoretical Physics, ETH Zurich, 8093 Z{\"u}rich, Switzerland\\
$^2$Institute of Quantum Electronics, ETH Zurich, 8093 Z{\"u}rich, Switzerland}
\end{center}

\section{Numerical methods}

\subsection{Spherical geometry and composite fermions}

We use the Haldane sphere for both exact diagonalization (ED) and the Jain
wave-function calculations~\cite{Haldane1983}.  A radial magnetic field generated by a monopole of strength $Q$ threads the sphere with
$N_\phi=2Q$ flux quanta.  The sphere has radius
\begin{equation}
 R=\sqrt{Q}\,\ell_B,
 \label{eq:sphere-radius}
\end{equation}
so its area is $\mathcal A = 4\pi R^2=2\pi N_\phi\ell_B^2$. The $\lambda$th Landau level ($\lambda=0,1,\ldots$) is an
angular-momentum multiplet with $j=Q+\lambda$ and
$m=-j,-j+1,\ldots,j$. It therefore contains
$2j+1=2Q+2\lambda+1$ orbitals. The Landau level orbitals, written in spinor coordinates, are
~\cite{WU1976,Wu1977}
\begin{equation}
\begin{aligned}
Y_{Q,Q+\lambda,m}(u,v)
={}&\mathcal N_{Q\lambda m}
\sum_{r=0}^{\lambda}(-1)^r
\binom{\lambda}{r}
\binom{2Q+\lambda}{Q+\lambda-m-r}
u^{Q+m+r}v^{Q-m+\lambda-r}\bar u^{\,r}\bar v^{\,\lambda-r},
\end{aligned}
\label{eq:full-monopole-harmonics}
\end{equation}
where $\mathcal N_{Q\lambda m}$ is the normalization.
Here, the spinor coordinates relate to the electron spherical coordinates by
\begin{equation}
 u_i=\cos\frac{\theta_i}{2}\,\e^{\ii\varphi_i/2},
 \qquad
 v_i=\sin\frac{\theta_i}{2}\,\e^{-\ii\varphi_i/2}.
 \label{eq:spinor-coordinates}
\end{equation}
Thus $(u,v)=(1,0)$ is the north pole. 

To analyze the various states on the sphere we use a composite fermion picture.
As discussed in the main text, this is achieved by attaching $2p$ vortices to each electron to form composite fermions, such that their effective monopole strength is
\begin{equation}
    2Q^* = N_\phi-2p(N_e-1).
\end{equation}
The Landau levels of the composite fermions (i.e., in the presence of $Q^*$) are referred to as $\Lambda$ levels. The number of orbitals in the first $n$ $\Lambda$ levels is
\begin{equation}
 N^{\Lambda}_0=\sum_{\lambda=0}^{n-1}(2Q^*+2\lambda+1)
 =n(2Q^*+n),
 \label{eq:closed-shell-count}
\end{equation}

A fractional quantum Hall state occurs when an integer number of $\Lambda$ levels are filled: $N_e =  N_0^{\Lambda}$. This implies that the physical parameters are related by
\begin{align}
    N_\phi = \frac{N_e}{\nu} - (n+2p),
\end{align}
where $\nu = n/(2np+1)$. Here the shift $\mathcal S= n + 2p$ is a topological quantum number on the sphere~\cite{Wen1992shift}.

A single anyon can be added to a fractional quantum Hall state by adding or removing a single composite fermion: $N_e =  N_0^{\Lambda} \pm 1$. In this case, the physical parameters are related by
\begin{align}
    N_\phi = \frac{N_e}{\nu} - (n+2p) \mp \frac{1}{n},
\end{align}
where the upper signs correspond to a quasiparticle, and the lower signs to a quasihole. For the quasihole, we choose the removed orbital from the highest occupied $\Lambda$ level. For the quasiparticle, the additional orbital is from the lowest unoccupied $\Lambda$ level.

We point out that the connection between the sphere and the disk geometries can be made near the north pole in the planar limit by taking $Q,N_e\rightarrow\infty$ at fixed
$\nu$. To achieve this, we set $z=r\e^{-\ii\varphi}$, with $r$ the chord distance on the sphere, and fix $l=Q-m$ to study orbitals near the north pole. Then applying a local gauge transformation gives
\begin{align}
    \frac{e^{-\ii Q\varphi}}{R}Y_{Q,Q,Q-l}
\propto z^l\left(1-\frac{|z|^2}{4R^2}\right)^{Q-l/2}
\xrightarrow[Q\to\infty]{}
z^l e^{-|z|^2/(4\ell_B^2)},
\end{align}
where we divide by \(R\) to convert the sphere’s solid-angle normalization to area normalization.
We thus observe that $l$ is the angular momentum on the disk as claimed in the main text.

\subsection{Exact diagonalization}

We project the Coulomb interaction and the impurity potential into the
physical LLL.  In the notation of the main text, the Hamiltonian is
\begin{equation}
 H=\sum_m U_m c_m^\dagger c_m
 +\frac{1}{2}\sum_{m_1,m_2,m_3,m_4}
 V_{m_1,m_2}^{m_3,m_4}
 c_{m_1}^\dagger c_{m_2}^\dagger c_{m_4}c_{m_3}.
 \label{eq:sm-ed-hamiltonian}
\end{equation}
Here $c_m^\dagger$ creates an electron in $Y_{Q,Q,m}$ and
$V_{m_1,m_2}^{m_3,m_4}$ are the spherical Coulomb matrix elements, whose
explicit form is given in Ref.~\cite{Fano1986configuration}.  The impurity
is placed at the north pole.  For any axially symmetric electron--impurity
potential $V(r)$, with chord distance
$r=2R\sin(\theta/2)$, its projected matrix elements are diagonal:
\begin{equation}
 U_m=\int d\Omega\,
 V\left(2R\sin\frac{\theta}{2}\right)|Y_{Q,Q,m}(\theta,\varphi)|^2.
 \label{eq:sm-impurity-matrix-element}
\end{equation}
For the Gaussian $V_a(r)=U\exp[-r^2/(2a^2)]$ used in the main text, this
becomes the one-dimensional integral
\begin{equation}
 U_m=U(2Q+1)\binom{2Q}{Q-m}
 \int_0^1 dx\,
 e^{-2R^2x/a^2}x^{Q-m}(1-x)^{Q+m},
 \label{eq:sm-gaussian-matrix-element}
\end{equation}
where $x=\sin^2(\theta/2)$.  This expression also makes clear that a narrow
impurity potential predominantly couples to the north-pole orbital $m=Q$.

We construct the occupation-number basis separately in each conserved
$L_z=\sum_m m n_m$ sector and diagonalize Eq.~\eqref{eq:sm-ed-hamiltonian}.
The axial impurity breaks conservation of total $L$ but preserves $L_z$;
the quoted ground-state energy is the minimum over all $L_z$ sectors.
Denoting the lowest energies of the reference fluid and one-anyon sector by
$E^{\rm FQH}(U)$ and $E^s(U)$, respectively, where
$s={\rm QH},{\rm QP}$, we use the main-text definition
\begin{equation}
E_{b;s}(U)=
\bigl[E^s(U)-E^{\rm FQH}(U)\bigr]
-\bigl[E^s(0)-E^{\rm FQH}(0)\bigr].
 \label{eq:sm-binding-energy}
\end{equation}
The second bracket subtracts the finite-size energy cost of changing sector,
leaving only the impurity-induced contribution.

For the contact results, $U_c$ denotes the coefficient of
$c_Q^\dagger c_Q$ after LLL projection, rather than the peak height $U$ of
the unprojected Gaussian.  Equation~\eqref{eq:sm-gaussian-matrix-element}
gives
\begin{equation}
 U_c=U_Q\simeq U\frac{2Q+1}{2Q}\frac{a^2}{\ell_B^2},
 \qquad a\ll\ell_B.
 \label{eq:sm-contact-strength}
\end{equation}
The contact limit therefore takes $a\rightarrow0$ at fixed $U_c$, and the
projected impurity becomes $U_c c_Q^\dagger c_Q$.

\subsection{Jain wave functions and Monte Carlo}

\subsubsection{Construction of Jain wave function}

The Jain wave functions follow from the composite fermion picture discussed above. In particular, the wave function reads
\begin{equation}
 \Psi_{n,p}=\PLLL\left[\Phi J^{2p}\right], \label{Eq:JainAgain}
\end{equation}
where $J^{2p}$ corresponds to the attachment of $2p$ vortices to each electron, $\Phi$ is the Slater determinant over occupied composite fermion orbitals and $\PLLL$ is the projection to the LLL. Here we write the Jastrow factor as $J=\prod_{i<j}D_{ij}$, where 
\begin{equation}
 D_{ij}=u_i v_j-u_j v_i.
 \label{eq:spherical-separation}
\end{equation}

In order to build $\Phi$, let $\mathcal O$ be the set of $N_e$
occupied composite-fermion orbitals at effective monopole strength $Q^*$.
The elements of $\mathcal O$ are $(\lambda_a,m_a)$, with $a=1,\ldots,N_e$. We then denote the Slater determinant by
\begin{equation}
 \Phi[\mathcal O]
 =\det_{1\leq a,i\leq N_e}
 \left[Y_{Q^*,Q^*+\lambda_a,m_a}(u_i,v_i)\right].
 \label{eq:sphere-slater}
\end{equation}

The FQH state fills the lowest $n$ $\Lambda$ levels at effective
monopole strength $Q^*$. Its occupied-orbital set is
\begin{equation}
 \mathcal O_{\rm FQH}
 =\{(\lambda,m): 0\leq\lambda<n,\;
       m=-Q^*-\lambda,\ldots,Q^*+\lambda\}.
\end{equation}
A quasihole is obtained by removing an orbital from the highest filled
$\Lambda$ level, while a quasiparticle occupies an orbital in the first
empty level. Using $l=Q^*-m$ to label that orbital, their sets are
\begin{align}
 \mathcal O_{{\rm QH},l}
 &=\mathcal O_{\rm FQH}\setminus\{(n-1,Q^*-l)\},\\
 \mathcal O_{{\rm QP},l}
 &=\mathcal O_{\rm FQH}\cup\{(n,Q^*-l)\}.
\end{align}

\subsubsection{Jain--Kamilla projection on the sphere}

Exact LLL projection of Eq.~\eqref{Eq:JainAgain} becomes
prohibitively expensive at large $N_e$.  We therefore use a well-established approximation known as the Jain--Kamilla (JK)
projection~\cite{JK} to simplify the numerics.  Define the single-particle Jastrow factors
\begin{equation}
 J_i=\prod_{j\ne i}D_{ij},
 \qquad
 \prod_iJ_i^p=(-1)^{pN_e(N_e-1)/2}J^{2p}.
 \label{eq:sphere-jk-identity}
\end{equation}
Suppressing the overall sign, this identity allows us to write the
unprojected Jain wave function as a determinant whose $i$th column
contains $J_i^p$:
\begin{equation}
\Phi[\mathcal O]J^{2p}
=\det_{1\leq a,i\leq N_e}
\left[Y_{Q^*,Q^*+\lambda_a,m_a}(u_i,v_i)J_i^p\right].
\end{equation}
The JK approximation projects each entry of this determinant separately.
Writing the projected entry as $J_i^p\widetilde Y_a(i)$, we obtain
\begin{align}
\widetilde Y_a(i)
&=J_i^{-p}\PLLL\left[
Y_{Q^*,Q^*+\lambda_a,m_a}(u_i,v_i)J_i^p
\right],
\label{eq:sphere-jk-orbital}\\
\Psi_{\rm JK}[\mathcal O]
&\propto\left(\prod_i J_i^p\right)\det_{1\leq a,i\leq N_e}
\left[\widetilde Y_a(i)\right].
\label{eq:sphere-jk-state}
\end{align}
Here, after moving conjugate spinors to the left, the projection to the LLL may be implemented, up to orbital normalization, by
\begin{equation}
 \bar u_i\longrightarrow\pdv{}{u_i},
 \qquad
 \bar v_i\longrightarrow\pdv{}{v_i}.
 \label{eq:sphere-projection-rule}
\end{equation}

\paragraph{Reduction to one monomial per orbital.}

The direct implementation of Eq.~\eqref{eq:sphere-jk-orbital} is also unnecessarily
expensive: the orbital in level $\lambda$ contains a sum of $\lambda+1$ terms, and the projection generates a separate sum through the general Leibniz rule for each term. We now show that both of these sums can be eliminated. At fixed $q$ and $m$, the orbitals have structure
\begin{equation}
 Y_{q,q+\lambda,m}\propto
 u^{q+m}v^{q-m}P_\lambda(x),
 \qquad x=|v|^2 = 1- |u|^2,
 \label{eq:monopole-polynomial-form}
\end{equation}
where $P_\lambda$ is a polynomial of degree $\lambda$.  The orbitals in
levels $0,\ldots,\lambda-1$ with the same $m$ span all allowed polynomials of
lower degree.  Since adding an occupied lower-level row to a row of a Slater
determinant does not change the determinant, only the highest-degree part of
$P_\lambda$ is relevant.  We may therefore choose the single-monomial
representatives
\begin{equation}
 Y_{q,q+\lambda,m}\ \widehat{=}\
 \begin{cases}
 \bar u^{\lambda}u^{q+m+\lambda}v^{q-m},&m\leq q,\\[3pt]
 \bar v^{\lambda}u^{q+m}v^{q+\lambda-m},&m>q,
 \end{cases}
 \label{eq:single-monomial-representatives}
\end{equation}
where $\widehat{=}$ denotes equality modulo occupied lower-$\Lambda$-level
rows and an irrelevant row normalization.

To see the second simplification, define
\begin{equation}
 F_k^u(i)=\frac{1}{J_i^p}\pdv[k]{J_i^p}{u_i},
 \qquad
 F_k^v(i)=\frac{1}{J_i^p}\pdv[k]{J_i^p}{v_i},
 \qquad F_0^{u,v}=1.
 \label{eq:sphere-F-functions}
\end{equation}
For the first line of Eq.~\eqref{eq:single-monomial-representatives}, let
$a=q+m+\lambda$ and $b=q-m$. Using the generalized Leibniz rule, the
projected JK element is proportional to
\begin{equation}
 \frac{1}{J_i^p}\pdv[\lambda]{u_i^a v_i^bJ_i^p}{u_i}
 =\sum_{k=0}^{\lambda}\binom{\lambda}{k}
 \frac{a!}{(a-\lambda+k)!}
 u_i^{a-\lambda+k}v_i^bF_k^u(i).
 \label{eq:jk-leibniz-sum}
\end{equation}
The term indexed by $k$ has monomial degree $2Q^*+k$ and
corresponds to a row from $\Lambda$ level $k$. Because every
lower level is completely filled, the terms with $k<\lambda_a$
can be eliminated by row operations. Only the $k=\lambda_a$
term remains. The same argument applies to the $\bar v$ rows.
Thus, for the $a$th occupied orbital $(\lambda_a,m_a)\in\mathcal O$,
the JK matrix entry is
\begin{equation}
 G_{ai}=
 \begin{cases}
 u_i^{Q^*+m_a+\lambda_a}v_i^{Q^*-m_a}F_{\lambda_a}^u(i),
   &m_a\leq Q^*,\\[3pt]
 u_i^{Q^*+m_a}v_i^{Q^*+\lambda_a-m_a}F_{\lambda_a}^v(i),
   &m_a>Q^*.
 \end{cases}
 \label{eq:sphere-jk-matrix}
\end{equation}
For any of the occupation sets $\mathcal O$ defined above, the
wave function sampled in Monte Carlo is
\begin{equation}
 \Psi_{\rm JK}[\mathcal O]\propto
 J^{2p}\det_{1\leq a,i\leq N_e}G_{ai}.
 \label{eq:sphere-jk-final}
\end{equation}

Equation~\eqref{eq:sphere-jk-matrix}
applies to the ground state, to a quasihole made in the highest filled level,
and to a quasiparticle added to the first empty level.  It would not apply
unchanged to a state with a hole in a lower $\Lambda$ level, because the row
needed to eliminate one of the Leibniz terms could then be absent.

\paragraph{Algebraic generation.}

The remaining functions $F_k^u$ and $F_k^v$ can be generated without repeatedly
differentiating $J_i^p$.  Define the power sums
\begin{align}
 S_r^u(i)&=\sum_{j\ne i}\frac{v_j^r}{D_{ij}^r},&
 S_r^v(i)&=(-1)^r\sum_{j\ne i}\frac{u_j^r}{D_{ij}^r}.
 \label{eq:sphere-power-sums}
\end{align}
The alternating sign in $S_r^v$ follows from
$\partial_{v_i}D_{ij}=-u_j$. Direct differentiation gives
\begin{equation}
 \partial_{u_i}S_r^u(i)=-rS_{r+1}^u(i),
 \qquad
 \partial_{v_i}S_r^v(i)=-rS_{r+1}^v(i).
 \label{eq:power-sum-derivative}
\end{equation}
By Eq.~\eqref{eq:sphere-F-functions},
$\partial_{u_i}^k J_i^p=J_i^pF_k^u(i)$. Differentiating once more and using
$\partial_{u_i}J_i^p=pS_1^u(i)J_i^p$ gives
$F_{k+1}^u=pS_1^uF_k^u+\partial_{u_i}F_k^u$; the $v_i$ derivative gives the
corresponding relation for $F_k^v$.
Let $\mathcal D$ act on polynomials in the $S_r$ by
$\mathcal D S_r=-rS_{r+1}$ and the product rule. Then both relations become
\begin{equation}
 F_0=1,
 \qquad
 F_{k+1}=pS_1F_k+\mathcal D F_k,
 \label{eq:F-recurrence}
\end{equation}
with $S_r=S_r^u$ for $F_k^u$ and $S_r=S_r^v$ for $F_k^v$. The first four
polynomials are
\begin{align}
 F_1={}&pS_1,\nonumber\\
 F_2={}&p^2S_1^2-pS_2,\nonumber\\
 F_3={}&p^3S_1^3-3p^2S_1S_2+2pS_3,\nonumber\\
 F_4={}&p^4S_1^4-6p^3S_1^2S_2+3p^2S_2^2
       +8p^2S_1S_3-6pS_4.
 \label{eq:first-F-polynomials}
\end{align}
These expressions are generated once and then evaluated numerically for each
particle.

\subsubsection{Density and impurity-overlap estimators}

The necessary expectation values are evaluated using the normalized probability density $|\Psi_{\rm JK}(X)|^2$, where $X=(\Omega_1,\ldots,\Omega_{N_e})$ are the electron positions on the sphere. The states that we consider all have definite $L_z$, and thus their one-body density is axially symmetric. The density profiles in Fig.~\ref{fig:Fig2} are obtained from a histogram of electron counts in equal-area bins of $\cos\theta$. In particular, for a bin $B$ of area $A_B$,
\begin{equation}
 \rho_B=\frac{1}{A_B}
 \left\langle\sum_{i=1}^{N_e}\mathbf{1}_{\Omega_i\in B}\right\rangle.
 \label{eq:density-histogram-estimator}
\end{equation}

For the anyon-impurity binding with a finite range Gaussian impurity, the overlap in Eq.~\eqref{eq:jain-binding-overlap} is constructed from the Monte Carlo samples via
\begin{equation}
 I_{s,l}(a)=
 \left\langle\sum_{i=1}^{N_e}e^{-r_i^2/(2a^2)}\right\rangle_{s,l}
 -\rho_{\rm ref} \int dA\,e^{-r^2/(2a^2)}.
 \label{eq:direct-overlap-estimator}
\end{equation}
On a sphere, the last integral is
$2\pi a^2[1-e^{-2R^2/a^2}]$.

\subsubsection{Direct estimator for the contact density}

The density exactly at the north pole is required for the contact results in
Table~\ref{tab:D}. However, the histogram method described above is poorly suited to this quantity: the result must be extrapolated to a cap of vanishing area where the density is typically rapidly varying. We instead determine this density by taking advantage of the fact that only one LLL level orbital has weight at the north pole.

Write $\mathrm N$ for the North-Pole position $(u,v)=(1,0)$, to distinguish
it from the electron number $N_e$. Let $\mathcal A=4\pi R^2$. The physical
LLL orbitals are the $\lambda=0$ monopole harmonics $Y_{Q,Q,m}$ of
Eq.~\eqref{eq:full-monopole-harmonics}. Relabeling them by $b=Q-m$ and
normalizing with respect to area $dA=R^2d\Omega$ gives
\begin{equation}
 \phi_b(u,v)=
 \left[\frac{2Q+1}{\mathcal A}\binom{2Q}{b}\right]^{1/2}
 u^{2Q-b}v^b,
 \qquad b=0,\ldots,2Q.
 \label{eq:physical-lll-basis}
\end{equation}
where $2Q=N_\phi$. Thus $\phi_b=Y_{Q,Q,Q-b}/R$ up to an orbital phase.
Only $b=0$ (equivalently, $m=Q$) is nonzero at $\mathrm N$, and
$|\phi_0(\mathrm N)|^2=(N_\phi+1)/\mathcal A$.  Therefore the point density
is fixed by the occupation $n_0=\langle c_Q^\dagger c_Q\rangle$ of this
single orbital:
\begin{equation}
 \rho({\rm N})=\frac{2Q+1}{\mathcal A}n_0.
 \label{eq:pole-density-occupation}
\end{equation}

To derive a direct estimator for $n_0$, fix the coordinates
$X_{\ne i}$ of all electrons except $i$ and expand the LLL many-body wave
function in the remaining coordinate:
\begin{equation}
 \Psi(\Omega_i,X_{\ne i})=
 \sum_{b=0}^{N_\phi}\phi_b(\Omega_i)A_b(X_{\ne i}).
 \label{eq:pole-lll-expansion}
\end{equation}
At the pole, all terms except $b=0$ vanish, so
$\Psi(\mathrm N_i,X_{\ne i})=\phi_0(\mathrm N)A_0(X_{\ne i})$.
If $P_0^{(i)}$ projects the $i$th electron onto $\phi_0$, it follows that
\begin{align}
 (P_0^{(i)}\Psi)(\Omega_i,X_{\ne i})
 &=\phi_0(\Omega_i)A_0(X_{\ne i})\nonumber\\
 &=u_i^{N_\phi}\Psi(\mathrm N_i,X_{\ne i}),
 \label{eq:pole-projector-action}
\end{align}
where we used
$\phi_0(\Omega_i)/\phi_0(\mathrm N)=u_i^{N_\phi}$.

The occupation is $n_0=\sum_i\langle P_0^{(i)}\rangle$.  Consequently, in
each configuration $X$ drawn with probability proportional to $|\Psi(X)|^2$,
we choose an electron label $I$ uniformly and measure
\begin{equation}
 \widehat n_0
 =N_e u_I^{N_\phi}
 \frac{\Psi(\mathrm N_I,X_{\ne I})}{\Psi(X)}.
 \label{eq:pinned-orbital-estimator}
\end{equation}
Indeed, writing $Z=\int dX\,|\Psi(X)|^2$, averaging this local estimator gives
\begin{equation}
 \left\langle\widehat n_0\right\rangle
 =\frac{1}{Z}\sum_{i=1}^{N_e}\int dX\,
 \Psi^*(X)(P_0^{(i)}\Psi)(X)=n_0.
 \label{eq:pole-estimator-proof}
\end{equation}
Here $\Psi(\mathrm N_I,X_{\ne I})$ is evaluated with electron $I$ at the
north pole and every other coordinate unchanged.  This temporary displacement
is made only for the extra wave-function evaluation; it is not proposed as a
move of the Markov chain.  Thus one additional wave-function evaluation per
measurement gives the density exactly at the pole, without a spatial
histogram.  The estimator can be complex on an individual configuration, but
its expectation value is real, so the imaginary part provides a useful
convergence check.

\subsubsection{Monte Carlo sampling and statistical errors}

The Monte Carlo sampling of $|\Psi_{\rm JK}|^2$ is implemented according
to the Metropolis algorithm. Each independent run begins with the
electrons placed uniformly at random on the sphere. The sequence of
configurations generated in one run is called a chain. We evolve each
chain through a sequence of sweeps, with $N_e$ attempted single-electron
moves per sweep. For each attempt, we select an electron at random,
choose a uniformly random tangent direction at its position, and
displace it along the sphere by a Gaussian-distributed angle. The
proposed configuration $X'$ is accepted with probability
$\min\{1,|\Psi_{\rm JK}(X')/\Psi_{\rm JK}(X)|^2\}$. During equilibration,
we tune the width of the angular proposal to target an acceptance rate
near $40\%$, then hold the width fixed during production.

The sampling effort is determined by the number of independent chains
and the numbers of equilibration and production sweeps per chain. For
the contact results in Table~\ref{tab:D}, the $n=2,3,4$ states use
16--32 chains per size, each with roughly $12{,}000$--$20{,}000$
production sweeps; the $n=1$ QP states use 96 chains with about
$33{,}000$ production sweeps. The $n=1$ QH pole density vanishes
exactly and needs no sampling. The large-system density profiles and
Gaussian overlaps use 96 chains of about $33{,}000$ production sweeps
per state. We check convergence of the pole estimator in
Eq.~\eqref{eq:pinned-orbital-estimator} using its imaginary part,
successive portions of each chain, and comparisons between independent
chains.

The errors on the observables estimated by Monte Carlo are calculated as follows. Consider a quantity $W$ that we aim to estimate (e.g., density). Let $W_c$ be the mean of chain $c$ and $\bar W=M^{-1}\sum_c W_c$. The reported sampling error is
\begin{equation}
 \sigma_W=\left[\frac{1}{M(M-1)}
 \sum_{c=1}^{M}(W_c-\bar W)^2\right]^{1/2}.
 \label{eq:chain-standard-error}
\end{equation}
This estimate assumes that the chains are independent and have
equilibrated. For the finite-size extrapolation another step is required to estimate the error, which we now discuss.

\subsubsection{Finite-size extrapolation of the contact depletion}

\begin{figure*}[t]
 \centering
 \includegraphics[width=0.92\textwidth]{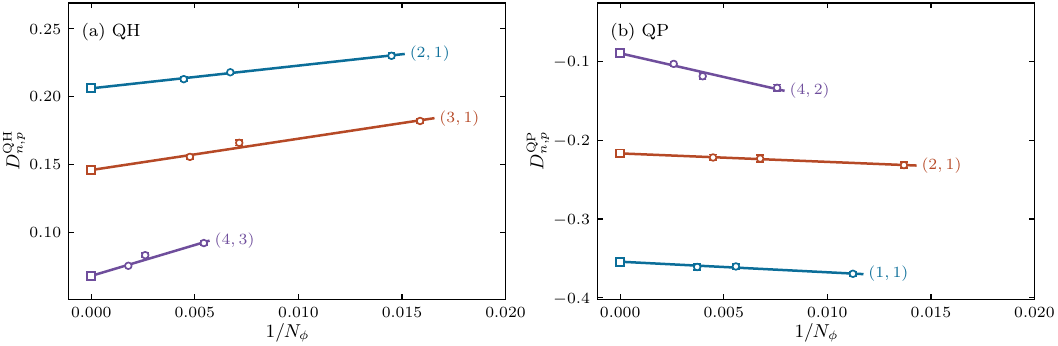}
 \caption{\textbf{Representative contact-depletion extrapolations for
 Table~\ref{tab:D}.} (a) Quasiholes and (b) quasiparticles. Open circles
 show the three finite-size Monte Carlo estimates with their uncertainties; solid lines are weighted fits linear in $1/N_\phi$, each labelled by
 its $(n,p)$, and open squares on $1/N_\phi=0$ mark the thermodynamic
 intercepts.}
 \label{fig:contact-extrapolation}
\end{figure*}

At each finite system size considered, the pole estimator gives $\rho({\rm N})=\rho_{s,l=0}(r=0)$. The finite-size depletion is 
\begin{equation}
 D_{n,p}^{s}(N_e)=2\pi\ell_B^2
 [\rho_{\rm ref}-\rho({\rm N})]
 =\nu\left(1+\frac{\mathcal S}{N_\phi}\right)
 -2\pi\ell_B^2\rho({\rm N}),
 \qquad \sigma_D=2\pi\ell_B^2\sigma_{\rho({\rm N})},
 \label{eq:finite-sphere-contact-conversion}
\end{equation}
where we have used
$\rho_{\rm ref}=\nu(N_\phi+\mathcal S)/\mathcal A$.
The entries in Table~\ref{tab:D} are the intercepts of linear fits
in $1/N_\phi$ to the Monte Carlo data at three system sizes,
$N_e\simeq30,60,90$. Each point is weighted by the inverse of its
variance. The stated uncertainty is the square root of the intercept
variance from the weighted-fit covariance matrix.
Figure~\ref{fig:contact-extrapolation} shows representative finite-size
data and extrapolations for quasiholes and quasiparticles.

\end{document}